\documentclass{aip-cp}

\usepackage[numbers]{natbib}
\usepackage{rotating}
\usepackage{graphicx}
\usepackage{upgreek}
\usepackage{amsmath}

\def\eg{{\it e.g.}}

\begin{document}

% Title portion
\title{Construction and Measurements of an Improved Vacuum-Swing-Adsorption Radon-Mitigation System}

\author{J.~Street\corref{cor1}}
\author{R.~Bunker}
\author{C.~Dunagan}
\author{X.~Loose}
\author{R.W.~Schnee}
\author{M.~Stark}
\author{K.~Sundarnath}
\author{D.~Tronstad}

\affil{Department of Physics, South Dakota School of Mines \& Technology, Rapid City, SD 57701}
\corresp[cor1]{Corresponding author: joseph.street@mines.sdsmt.edu}
\maketitle

\begin{abstract}
In order to reduce backgrounds from radon-daughter plate-out onto detector surfaces, an ultra-low-radon cleanroom is being commissioned at the South Dakota School of Mines and Technology. An improved vacuum-swing-adsorption radon mitigation system and cleanroom build upon a previous design implemented at Syracuse University that achieved radon levels of $\sim$0.2\,Bq\,m$^{-3}$. This improved system will employ a better pump and larger carbon beds feeding a redesigned cleanroom with an internal HVAC unit and aged water for humidification. With the rebuilt (original) radon mitigation system, the new low-radon cleanroom has already achieved a $>$\,300$\times$ reduction from an input activity of $58.6\pm0.7$\,Bq\,m$^{-3}$ to a cleanroom activity of $0.13\pm0.06$\,Bq\,m$^{-3}$.
% over a six day run.   (not needed but could put back)
\end{abstract}

% Head 1
\section{INTRODUCTION TO RADON MITIGATION}
A potential source of dominant backgrounds for many rare-event searches or screening detectors is from radon daughter plate-out~\citep{LRT2013simgen}. 
Backgrounds from 
$^{210}$Pb 
%(a beta emitter) 
and the recoiling $^{206}$Pb nucleus from the $\alpha$ decay of $^{210}$Po 
are the dominant 
%low-mass WIMP 
low-energy
backgrounds for XMASS~\citep{LRT2015xmass}, SuperCDMS Soudan~\citep{supercdms_lowmass2014} and EDELWEISS~\citep{LRT2013edelweiss},
and are expected to be dominant for SuperCDMS SNOLAB and the BetaCage screener~\citep{LRT2013bunker,LRT2015schnee} without improvements.
Radon daughters on surfaces may dominate for SuperNEMO~\citep{LRT2013SuperNEMO}
and CUORE~\citep{LRT2015cuore}.
Both neutrons from ($\alpha,n$) reactions and $^{206}$Pb  recoils are important for LZ~\citep{lux2014backgrounds},  XENON1T, and DArKSIDE.
Storing and assembling detector components may be possible using vacuum glove boxes or by cleaning components upon commissioning (\eg~\citep{LRT2013schneeEP}). However, when the components are large or require delicate assembly, a vacuum glove box can be impractical. Cleaning can also be impractical when the components are delicate or have complex geometries.

To create a radon-mitigated, breathable-air environment one may consider two system classes: continuous flow through a filtration column and swing flow through two or more filtration columns. The filtration columns are usually filled with activated carbon. The continuous flow system (\eg~\citep{nemoLRT2006}) operates on the basis that some considerable fraction of radon decays before exiting the column. For an ideal column, the final radon concentration $C_\text{final}=C_\text{initial}\exp{\left(-t/t_\text{Rn}\right)}$, where $C_\text{initial}$ is the radon concentration of the input air, $t$ is the characteristic break-through time of the filter, and $t_\text{Rn}=5.38$\,days is the mean lifetime of radon. To increase the break-through time, and therefore make a continuous flow system practical, one must cool the carbon to reduce desorption of radon. Continuous flow systems are relatively simple and robust, are commercially available, and typically achieve reduction factors of $\sim$1000$\times$, to $\sim$10--30\,mBq\,m$^{-3}$. 

\begin{figure}
	\begin{tabular}{c c}

(a)\includegraphics[width=0.49\linewidth]{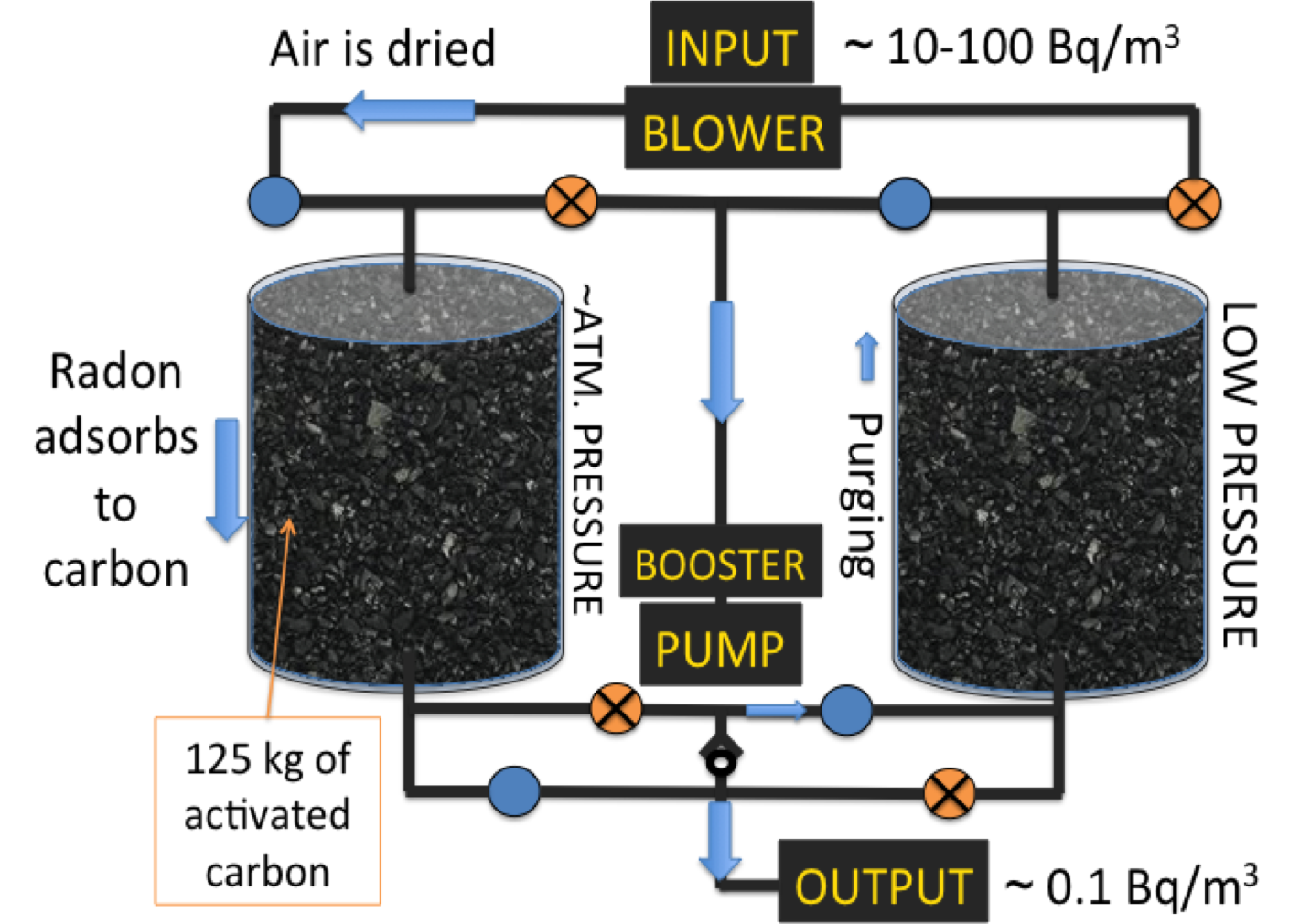} &(b)\includegraphics[width=0.45\textwidth]{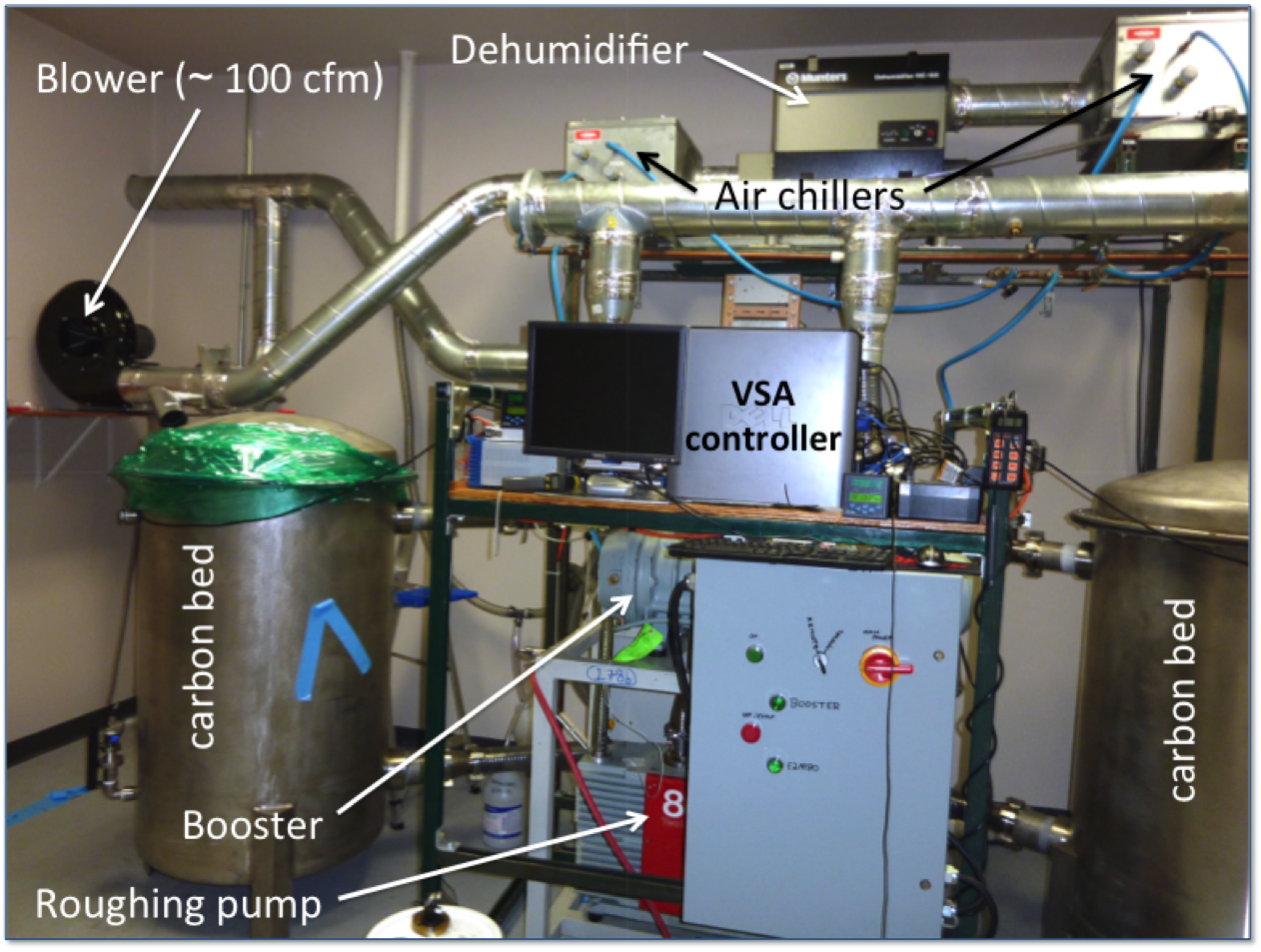}

	\end{tabular}
 	\caption{Schematic diagram $(a)$ and photo $(b)$ of the SDSM\&T VSA system.  	High-radon air is brought in either from outdoors ($\sim$10\,Bq m$^{-3}$) or from the building ($\sim$100\,Bq m$^{-3}$). The air is dried and passed through one of the two carbon columns (open circles denote open valves, while $\otimes$ denotes a closed valve). The radon in the air adsorbs preferentially to the carbon (compared to N$_2$ or O$_2$). The radon slowly migrates through the column, adsorbing and desorbing,	while low-radon air flows to the cleanroom. A small amount of this low-radon air flows back into the other carbon column, while it is 	pumped to a few Torr, helping radon to desorb from the carbon and be exhausted. Before any radon breaks through the left column, the system swings and the cycle repeats. 
	}
 	\label{VSAdiagram} 
\end{figure}

In a swing flow system, two or more filtration columns are used. While filtering through one column the other is regenerated using either low pressure or high temperatures to allow radon to desorb efficiently and be exhausted. For a vacuum-swing-adsorption (VSA) system (\eg~\citep{LRT2004Pocar,PocarThesis, LRT2013schneeVSA}), high-radon input air is filtered through the first column while the second column is pumped down to $\sim$1\,Torr. Well before the break-through time, the path of the high-radon input air is switched so that it flows through the second column instead, allowing the first to regenerate. For an ideal column, no radon reaches the output. Swing flow systems are more complicated both in operation and analysis. A VSA system ($e.g.$ Fig.~\ref{VSAdiagram}) can potentially outperform a continuous flow system at a lower cost. Temperature-swing systems (\eg~\citep{LRT2010HallinRadon}) should provide best performance but at the highest cost and complexity. We use the VSA technique because of its lower cost and potential for excellent radon reduction.

%A vacuum-swing adsorption system takes advantage of the filtering medium's greater adsorption capacity at high pressures. The carbon is regenerated by flowing a small fraction $f$ of filtered gas of mass flow $F$ back through the tank at low purge pressure $P_\text{purge}$. The volume purge flow
%\begin{equation}
%\phi_{\text{purge}} = \frac{P_\text{atm}} {P_\text{purge}} f F  = \frac{P_\text{atm}} {P_\text{purge}} f \phi_{\text{feed}}  .
%\end{equation}
%On each cycle, the radon front is pushed back more than it moves forward if  the volume flow gain $G \equiv \phi_\text{purge} / \phi_\text{feed}>1$, that is if $f P_\text{atm} > P_\text{purge}$.

%\section{DESIGN AND CONSTRUCTION OF THE SDSM\&T RADON-MITIGATION SYSTEM}
\section{THE SDSM\&T RADON-MITIGATION SYSTEM AND CLEANROOM}
The system built at the South Dakota School of Mines \& Technology (SDSM\&T) consists of two parts: a VSA radon filter and a low-radon cleanroom. Figure~\ref{VSAdiagram} shows the VSA rebuilt as it was at Syracuse University~\cite{LRT2013schneeVSA} (based on the design in~\citep{LRT2004Pocar,PocarThesis}) with a few modifications due to the geometry of its new living space. 
The system has a blower that takes in air from either outdoors or inside the lab  at a 100\,cfm capacity. The input air is dried with a dehumidifier and passes through chillers to maintain air temperature. Each column is filled with 125\,kg activated carbon. With the use of a booster to accelerate pumping speed at low pressure, a column being purged is evacuated to $<$2.5\,Torr in 7\,min. High-radon air is filtered through one column for 40\,min while the other column is regenerated, giving a full swing-cycle period of 80\,min.

The low-radon cleanroom is an evolution of the Syracuse University cleanroom design. 
The cleanroom was 
built with materials having sufficiently low emanation and permeation of radon. The area of the cleanroom is 21\,$\times$\,9.25\,ft$^2$ with an additional 5\,$\times$\,9.25\,ft$^2$ of internal anteroom space and a ceiling height of 8\,ft. Recent measurements indicate that the cleanroom is class 100 when empty.

The cleanroom was constructed primarily of aluminum. All seams were sealed with butyl rubber and aluminum tape making the cleanroom very leak-tight and easily over-pressured. Although the VSA system can provide low-radon air at 100\,cfm, the cleanroom can be over-pressured effectively (to $\sim$0.25 inches of water) with only 15\,cfm of make-up air. The windows are made of $\geq$ 1/8 inch polycarbonate, contributing $\leq8$\,mBq\,m$^{-3}$ of $^{222}$Rn activity to the cleanroom. If diffusion through the windows becomes the dominant source of radon (following the upgrades described below), they will be covered with metal and/or thickened. Low-radon air is delivered to the cleanroom from the VSA system via galvanized-steel, 22-gauge spiral ducting.

A notable feature of this cleanroom is the internal placement of the HVAC. During operation, the HVAC circulates low-radon air through HEPA filters and then back into the cleanroom at $\sim$1000\,cfm. A negative pressure region forms behind the blower
%, which drives the circulation, 
inside the HVAC. Placing the HVAC outside the cleanroom presents the difficulty of sealing it sufficiently to prevent high-radon air from leaking into the circulation path. 
Leaks of radon into the HVAC limited the system at Syracuse.
%This placement issue was a limiting factor for Syracuse system. 
Situating the HVAC inside the cleanroom mitigates this issue.

\begin{figure}[t]
		\begin{tabular}{c c}

(a)\includegraphics[width=0.48\textwidth]{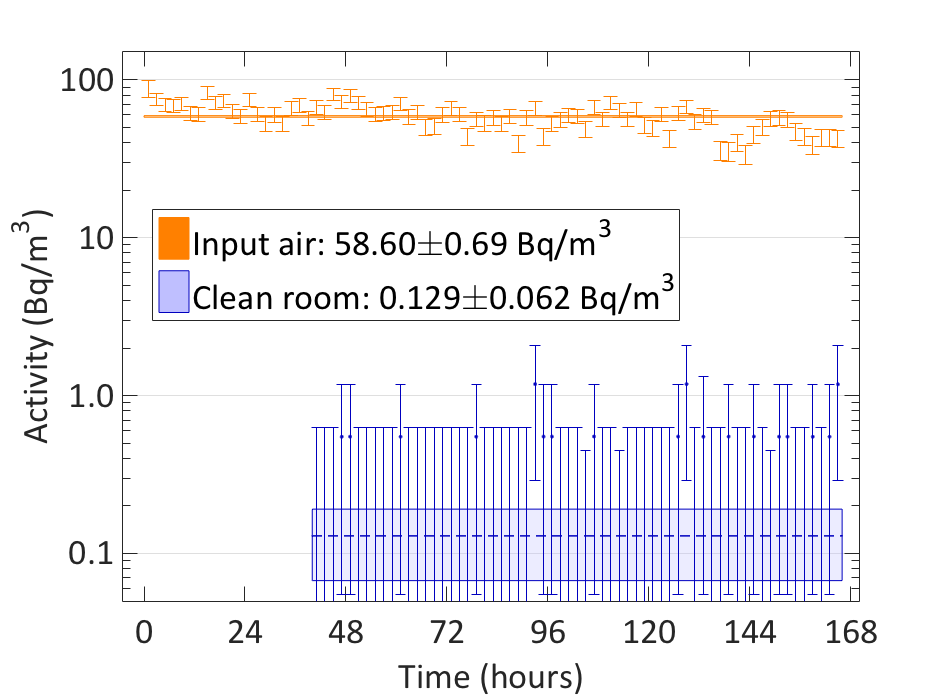}&(b)\includegraphics[width=0.48\textwidth]{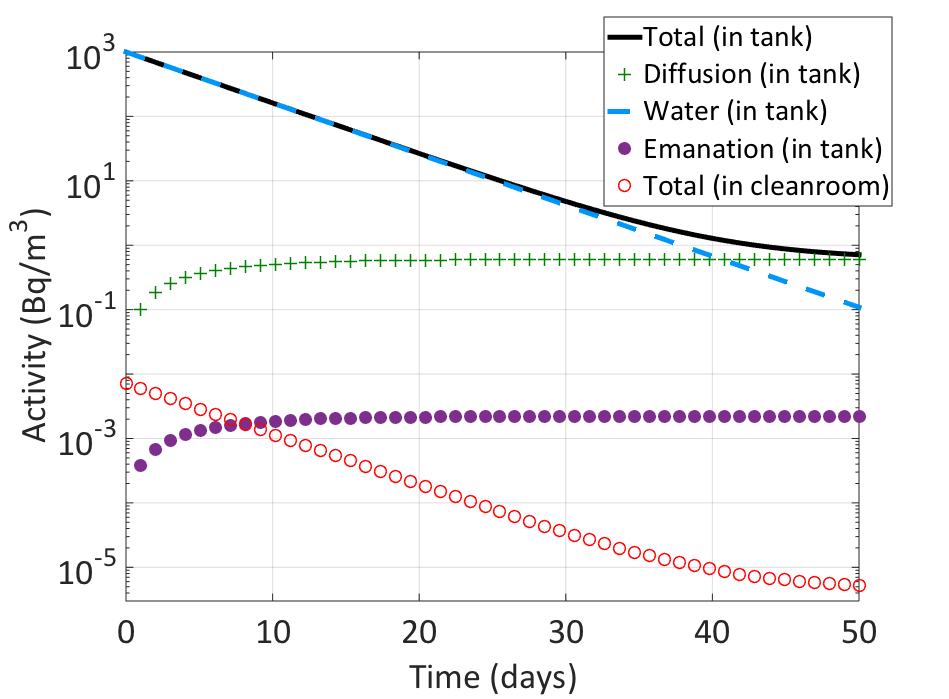}

		\end{tabular}
		\caption{$(a)$ Radon activity of the input air (higher, orange error bars with 7-day average) compared to that of the air in the cleanroom (lower, blue error bars with 5-day average), showing a $>$\,300$\times$ reduction in activity from $58.6\pm0.7$\,Bq\,m$^{-3}$ to $0.13\pm0.06$\,Bq\,m$^{-3}$.
		 $(b)$ Expected contributions from the initial water (blue dashes), from diffusion (green $+$'s), and from emanation (purple dots) to the total radon activity in the water aging tank (black solid).  Radon emanation is  negligible compared to radon diffusion, which limits this system for aging times longer than about 40 days. 
		 The corresponding contribution to the radon activity in the cleanroom (red circles) is diluted by over 5 orders of magnitude. }
 		\label{VSAresults} 
\end{figure}

The VSA system and cleanroom, commissioned at SDSM\&T, recently achieved a radon reduction of $>$\,300$\times$ to $0.13\pm0.06$\,Bq\,m$^{-3}$, as shown in Fig.~\ref{VSAresults}a. %(Fig. 2, $(b)$). 
%The input air was drawn from the lab with a mean activity of $58.60\pm0.69$\,Bq\,m$^{-3}$. 
About 100\,cfm of air was drawn from the lab and conditioned to a temperature of $\sim$16\,C and dew point of $-$17\,C. A fraction of this air was used as input to the VSA system, resulting in 20\,cfm of low-radon make-air for the cleanroom.
%. The low-radon make-up air of 20\,cfm and cleanroom recirculation rate was $\sim$1000\,cfm produced $>$1/4 inches of water overpressure. 
This result is consistent with the performance of the system while at Syracuse but has the added benefit (because the input air has higher activity than at Syracuse) of demonstrating the greater activity reduction of which this system is capable. The SDSM\&T system indicates that the VSA technique is a viable low-cost alternative to continuous flow systems.

\subsection{Low-Radon Humidification System and Other Planned Upgrades }
%Low humidity environments allow charge to build to levels high enough to damage electrical devices, start fires, and are uncomfortable to occupants. A cleanroom using a  vacuum-swing-adsorption (VSA) 
The VSA radon mitigation system tends to reduce relative humidity of the cleanroom air to $\sim$5\%. Using water for humidification without reducing its activity would introduce radon into the cleanroom.
To reduce water activity, we will age the water until a sufficient fraction of radon has decayed away.

The water to be used is expected to have an activity of $\sim$1000\,Bq\,m$^{-3}$. If no aging is done and the water is used for increasing the relative humidity in the cleanroom, the activity contribution to the cleanroom would be $\sim$6\,mBq\,m$^{-3}$ because the vapor density of water at 35\% relative humidity is $\sim$$6\times10^{-6}$\,g\,cm$^{-3}$. As shown in Fig.~\ref{VSAresults}b, after aging, the activity will be reduced by 100$\times$ and will contribute only 100's of $\upmu$Bq\,m$^{-3}$. 

The low-radon humidification system consists of two large (110\,gal) tanks made of low-density polyethylene (LDPE).
Water from a local reverse osmosis system fills the first tank. The water is aged until its activity is sufficiently low (about 10\,Bq\,m$^{-3}$) and then transferred to the second tank. Upon transferring, the first tank is again filled and the aging process repeats. The second tank supplies low-radon water to the cleanroom humidification system at a rate of about 3\,gal\,day$^{-1}$. The system is designed such that the aged water in the second tank 
lasts longer than the aging time. In addition, an in-line filter will  prevent any particulates from entering the cleanroom.

Sufficient aging time is determined by the intrinsic activity of the water, with consideration of radon emanation from the LDPE tank, and radon diffusion through the tank walls, as shown in Fig.~\ref{VSAresults}b.
Aging for 35\,days is sufficient to contribute only 100's of $\upmu$Bq\,m$^{-3}$ to the cleanroom. Further aging makes little difference as the total activity asymptotically approaches the contribution from diffusion, which limits this system.
If a lower activity is needed in the future, switching to metal tanks would reduce the contribution from humidification to the order of 10\,nBq\,m$^{-3}$, because LDPE has a diffusion coefficient on order of 10$^{-11}$\,m$^2$\,s$^{-1}$ while metal is $\sim$10$^{-15}$\,m$^2$\,s$^{-1}$~\cite{RadonDiffusionJiranek}.

Other radon-mitigation upgrades that have yet to be installed are taller carbon columns (providing an additional 175\,kg activated carbon each) and a more powerful roughing pump, which should allow easier maintenance as well as providing a small improvement in performance. In principle, the increased height of the carbon columns should provide additional reduction in radon concentration of another factor $\sim$100--1000$\times$, potentially providing world-leading radon reduction at a modest cost.  

% Acknowledgement
\section{ACKNOWLEDGMENTS}
This work was supported in part by the National Science Foundation (Grants No. PHY-1205898, PHY-1506633, and PHY-1546843), the Department of Energy (Grant No. DE-AC02-05CH1123), and the state of South Dakota.

% References

%\nocite{*}
%\bibliographystyle{aipnum-cp}
%\bibliographystyle{aipproc}
%\bibliography{schnee}

\begin{thebibliography}{15}
\expandafter\ifx\csname natexlab\endcsname\relax\def\natexlab#1{#1}\fi
\providecommand{\enquote}[1]{``#1''}
\expandafter\ifx\csname url\endcsname\relax
  \def\url#1{\texttt{#1}}\fi
\expandafter\ifx\csname urlprefix\endcsname\relax\def\urlprefix{URL }\fi
\providecommand{\eprint}[2][]{\url{#2}}

\bibitem[{Simgen}(2013)]{LRT2013simgen}
H.~{Simgen}, \enquote{{Radon assay and purification techniques},} in
  \emph{Topical Workshop on Low Radioactivity Techniques: LRT 2013}, edited by
  L.~{Miramonti}, and L.~{Pandola}, 2013, vol. 1549 of \emph{American Institute
  of Physics Conference Series}, pp. 102--107.

\bibitem[{Kobayashi}(2015)]{LRT2015xmass}
K.~{Kobayashi}, \enquote{{Surface purity control during XMASS detector
  refurbishment},} in \emph{Topical Workshop on Low Radioactivity Techniques:
  LRT 2015}, edited by J.~L. {Orrell}, American Institute of Physics Conference
  Series, 2015,
  {https://www.npl.washington.edu/indico/contributionDisplay.py?contribId=24\&%
confId=5}.

\bibitem[{Agnese} et~al.(2014)]{supercdms_lowmass2014}
R.~{Agnese}  {\it et al.}  (SuperCDMS Collaboration),
 \emph{Physical Review Letters}
  \textbf{112}, 241302 (2014), \eprint{1402.7137}.

\bibitem[{Navick} and {EDELWEISS Collaboration}(2013)]{LRT2013edelweiss}
X.-F. {Navick}, and {EDELWEISS Collaboration}, \enquote{{Background suppression
  in the Edelweiss-III experiment},} in \emph{Topical Workshop on Low
  Radioactivity Techniques: LRT 2013}, edited by L.~{Miramonti}, and
  L.~{Pandola}, 2013, vol. 1549 of \emph{American Institute of Physics
  Conference Series}, pp. 148--151.

\bibitem[{Bunker} et~al.(2013)]{LRT2013bunker}
R.~{Bunker}, Z.~{Ahmed}, M.~A. {Bowles}, S.~R. {Golwala}, D.~R. {Grant},
  M.~{Kos}, R.~H. {Nelson}, R.~W. {Schnee}, A.~{Rider}, B.~{Wang}, and
  A.~{Zahn}, \enquote{{The BetaCage, an ultra-sensitive screener for surface
  contamination},} in \emph{Topical Workshop on Low Radioactivity Techniques:
  LRT 2013}, edited by L.~{Miramonti}, and L.~{Pandola}, 2013, vol. 1549 of
  \emph{American Institute of Physics Conference Series}, pp. 132--135.

\bibitem[{Schnee}(2015)]{LRT2015schnee}
R.~W. {Schnee}, {The BetaCage, an Ultra-sensitive Screener for Surface
  Contamination}, {Talk at the Topical Workshop on Low Radioactivity
  Techniques: LRT 2015} (2015),
  {https://www.npl.washington.edu/indico/contributionDisplay.py?contribId=48\&%
confId=5}.

\bibitem[{Mott} and {SuperNEMO Collaboration}(2013)]{LRT2013SuperNEMO}
J.~{Mott}, and {SuperNEMO Collaboration}, \enquote{{Low-background tracker
  development for SuperNEMO},} in \emph{Topical Workshop on Low Radioactivity
  Techniques: LRT 2013}, edited by L.~{Miramonti}, and L.~{Pandola}, 2013, vol.
  1549 of \emph{American Institute of Physics Conference Series}, pp. 152--155.

\bibitem[{Wang}(2015)]{LRT2015cuore}
B.~{Wang}, \enquote{{Background analysis techniques in the CUORE experiment},}
  in \emph{Topical Workshop on Low Radioactivity Techniques: LRT 2015}, edited
  by J.~L. {Orrell}, American Institute of Physics Conference Series, 2015,
  {https://www.npl.washington.edu/indico/contributionDisplay.py?contribId=82\&%
confId=5}.

\bibitem[{Akerib} et~al.(2014)]{lux2014backgrounds}
D.~S. {Akerib} {\it et al.}  (LUX Collaboration),
  \emph{Astroparticle Physics} \textbf{62}, 33--46 (2014), \eprint{1403.1299}.

\bibitem[{Schnee} et~al.(2013{\natexlab{a}})]{LRT2013schneeEP}
R.~W. {Schnee}, M.~A. {Bowles}, R.~{Bunker}, K.~{McCabe}, J.~{White},
  P.~{Cushman}, M.~{Pepin}, and V.~E. {Guiseppe}, \enquote{{Removal of
  long-lived $^{222}$Rn daughters by electropolishing thin layers of stainless
  steel},} in \emph{Topical Workshop on Low Radioactivity Techniques: LRT 201},
  edited by L.~{Miramonti}, and L.~{Pandola}, 2013{\natexlab{a}}, vol. 1549 of
  \emph{American Institute of Physics Conference Series}, pp. 128--131.

\bibitem[{Nachab}(2007)]{nemoLRT2006}
A.~{Nachab}, \enquote{{Radon reduction and radon monitoring in the NEMO
  experiment},} in \emph{AIP Conf. Proc. 897: Topical Workshop on Low
  Radioactivity Techniques: LRT 2006}, edited by P.~{Loaiza}, American
  Institute of Physics, Melville, NY, 2007, pp. 35--39.

\bibitem[{Pocar}(2005)]{LRT2004Pocar}
A.~{Pocar}, \enquote{{Low background techniques for the Borexino nylon
  vessels},} in \emph{Topical Workshop on Low Radioactivity Techniques: LRT
  2004.}, edited by B.~{Cleveland}, R.~{Ford}, and M.~{Chen}, 2005, vol. 785 of
  \emph{American Institute of Physics Conference Series}, pp. 153--162,
  \eprint{arXiv:physics/0503243}.

\bibitem[{Pocar}(2003)]{PocarThesis}
A.~{Pocar}, \emph{{Low Background Techniques and Experimental Challenges for
  Borexino and its Nylon Vessels}}, Ph.D. thesis, Princeton University (2003).

\bibitem[{Schnee} et~al.(2013{\natexlab{b}})]{LRT2013schneeVSA}
R.~W. {Schnee}, R.~{Bunker}, G.~{Ghulam}, D.~{Jardin}, M.~{Kos}, and A.~S.
  {Tenney}, \enquote{{Construction and measurements of a
  vacuum-swing-adsorption radon-mitigation system},} in \emph{Topical Workshop
  on Low Radioactivity Techniques: LRT 2013}, edited by L.~{Miramonti}, and
  L.~{Pandola}, 2013{\natexlab{b}}, vol. 1549 of \emph{American Institute of
  Physics Conference Series}, pp. 116--119.

\bibitem[{Grant} et~al.(2011)]{LRT2010HallinRadon}
D.~{Grant}, A.~{Hallin}, S.~{Hanchurak}, C.~{Krauss}, S.~{Liu}, and R.~{Soluk},
  \enquote{{Low Radon Cleanroom at the University of Alberta},} in
  \emph{Topical Workshop on Low Radioactivity Techniques: LRT 2010}, edited by
  R.~{Ford}, 2011, vol. 1338 of \emph{American Institute of Physics Conference
  Series}, pp. 161--163.

\bibitem[{Jiranek} et~al.(2011)]{RadonDiffusionJiranek}
M.~{Jiranek}, and M.~{Kotrbata}
  \emph{Radiation Protection Dosimetry} \textbf{145}, 178--183 (2011).

\end{thebibliography}

\end{document}